\documentstyle[preprint,aps]{revtex}
\begin{document}  

\title {  A Phase-Field Model of Spiral Dendritic Growth } 
\author {Royce Kam and Herbert Levine}
\address
{Department of Physics\\
and\\
Institute for Nonlinear Science\\
University of California, San Diego\\
La Jolla, CA 92093-0402\\
}
\maketitle

\begin{abstract} 
Domains of condensed-phase monolayers of chiral molecules exhibit a
variety of interesting nonequilibrium structures when formed via pressurization.
To model these domain patterns, we add a complex field  describing the tilt 
degree of freedom to an (anisotropic) complex-phase-field solidification 
model.  The resulting formalism allows for the inclusion
of (in general, non-reflection symmetric) interactions between the tilt, the 
solid-liquid interface, and the bond orientation.  Simulations demonstrate 
the ability of the model to exhibit spiral dendritic growth.  
\end{abstract} 

\section*{I.  Introduction}
Several experiments performed over the last decade have explored the myriad
complex domain shapes formed during the growth of condensed-phase 
Langmuir monolayers~\cite{review}.  An interesting case is the pressure-induced 
growth of condensed-phases whose constituent molecules are chiral.  Here the growth 
shapes range from simple spiral dendrites to intricate chiral
fractal structures~\cite{Weis,Yang,Heckl}.  
More recently, experiments with racemic monolayers have exhibited initially achiral 
dense-branched growth, followed by the addition of chiral ``hooks'' at the edges 
of the initial domains~\cite{Stine}.  

In the same decade, the invention of the phase-field model provided
a means for studying not only these isolated growth elements
(such as dendrites), but also for elucidating the character of the global
morphology of these domain growths.
In a standard phase-field model of solidification, the phase of the material 
is represented by a real order parameter field whose time evolution is coupled 
to a diffusive field~\cite{Collins,Langer,Caginalp,Penrose}.  As the liquid 
solidifies, the latent heat released by the advancing solid front acts as a 
source to the diffusion field.  In the sharp-interface limit,
this formulation recovers the traditional free-boundary solidification
equations. Simulations of this model are able to reproduce the dense-branched 
structures seen in experiments with isotropic materials.  In addition, 
anisotropic terms may be added to the standard phase-field model allowing for 
simulation of the stable dendrites seen in experiments with anisotropic 
materials~\cite{Kupferman,McFadden,Karma}.  Still, the chiral dendrites
seen in experiments remain beyond the grasp of these models.

In a recent advance\cite{Kam}, the standard phase-field model was 
generalized to include information regarding the orientational order 
of the ``solid" phase (for example, the orientation of the crystal
axes of a solid) by incorporating this information directly into 
the phase of a complex order parameter.  Prior to this work, the 
implicit ``crystal axes" were fixed with respect to the coordinate
axes.  While the introduction of a complex order parameter allowed the ``crystal axes" to 
bend dynamically, chiral structures were still excluded by this model.  
Even with the inclusion of non-reflection symmetric anisotropy 
(a prerequisite for chiral growth) the dendrites failed to 
spiral.  This anisotropy merely lead to the growth of non-reflection
symmetric dendrites growing at constant velocity\cite{Brener}.

Some clues as to the physical origin of spiral dendrites have been provided 
by equilibrium studies of textures in Langmuir monolayers (single layers
of amphiphillic molecules at an air-water
interface)~\cite{Fischer,Selinger}.  
In these studies, many of the condensed phases are observed to have hexatic
orientational order.  In one, called the ``$LS$-phase", the average
direction of the hydrophobic tails is perpendicular to the air-water
interface.  In others, the average  tail alignment has a component in
the plane of the air-water interface.  Landau free energy descriptions
of the interactions between the tilt and the bond orientation have been
formulated which successfully describe the phase transitions among the
various condensed phases\cite{Selinger,Kaganer}.  

With this motivation, our strategy will be to add a
complex order parameter representing the tilt to an
anisotropic complex-phase-field model, and to couple the
tilt direction both to the bond orientation and to the
orientation of the solid-liquid interface.  As we shall see, the 
resulting model is indeed capable of supporting the growth of
spiral dendritic structures.  We interpret this as proof that 
these patterns can be understood as being the result of the
inconsistency of the tilt preferring to point both along a particular
direction with respect to the interface normal and along a particular
direction with respect to the crystalline anisotropy.  That is, we conclude
that spiral dendrites can arise from frustration of the tilt field.  

\section*{II.  The Model}
A complex-phase-field model describing diffusion-limited solidification is 
defined by the equations~\cite{Kam}:
\begin{eqnarray} -\Gamma \frac{d \Psi}{dt} &=& \frac{\delta F[\Psi,U]}{\delta \Psi^*} 
\label{PsiEqn}
\end{eqnarray} 
\begin{eqnarray}
D\nabla^2 U - \frac{dU}{dt} &=& -\frac{d |\Psi|^2}{dt}
\label{UEqn}
\end{eqnarray}
\begin{eqnarray}
F &=& F_{\Psi} \equiv \int d{\vec r} \ \kappa^2 |\nabla \Psi|^2 +  |\Psi|^2 - 
2|\Psi|^4 +  |\Psi| ^6 +  T |\Psi| ^2 \mbox{tanh}(\lambda U) 
\label{DoubleWell}
\end{eqnarray}
where $\Psi \equiv |\Psi| e^{iN\theta}$ is an N-fold rotationally 
symmetric complex order parameter whose argument indicates the orientation 
of the crystal axes with respect to the x-axis.  The first term in 
equation (\ref{DoubleWell}) is an energy cost for spatial inhomogeneity, which 
gives rise to a surface tension proportional to $\kappa$.  The remainder of 
equation (\ref{DoubleWell}) represents a ``double-well'' whose minima 
at $|\Psi|=0$ (liquid) and $|\Psi|=1$ (solid) are tilted by coupling to 
the diffusion field $U$ so as to favor the appropriate phase.

Physically, the bond orientational order leads to an anisotropy in
the surface tension.  Previous workers have included these 
anisotropic effects in various ways~\cite{Caginalp,Kupferman,Kam}.
Kobayashi\cite{Kobayashi} and Wheeler, Murray and Schaefer\cite{Wheeler} 
include anisotropy by allowing the surface tension coefficient 
to depend on the local orientation
of the gradient of the order parameter.  McFadden, et. al. \cite{McFadden}
later provided a thermodynamic formulation of this method, as well as 
an asymptotic analysis of the sharp-interface limit.  We adopt this 
method of including anisotropy, defining $\kappa$
as a function of both the crystal axis orientation and the normal to the interface.
  For an N-fold anisotropy we define 
\begin{eqnarray}
\kappa(\Psi) = \kappa_0 \left\{ 1 + \eta_\kappa |\Psi|^2  | \nabla |\Psi|^2|^2 \cos{N (\theta_n-\theta)} \right\},
\label{AnisotropicKappa}
\end{eqnarray}
with the normal to the interface (pointing from the solid into the liquid) 
defined, $\theta_n \equiv\tan^{-1}{(\partial_y |\Psi|^2/\partial_x|\Psi|^2)}$.  
The resulting dendrites will grow along the N-fold crystal axes.

To describe the ``tilt'' of the molecules, we define a new complex order 
parameter $\Phi \equiv |\Phi| e^{i \phi}$ whose magnitude reflects the local 
degree of tail alignment and whose argument indicates the direction of tilt in the 
plane of the air-water interface.  
We again assume the time-dependence
\begin{eqnarray} -\Gamma_\phi \frac{d \Phi}{dt} &=& \frac{\delta F[\Psi,\Phi,U]}{\delta \Phi^*} .
\label{PhiEqn}
\end{eqnarray} 
Experimentally, only liquid-condensed phases exhibit tilt.  To incorporate this fact, consider the energy
\begin{eqnarray} F_{mag} &=& \eta_{mag} \int{d{\vec r} \left( |\Phi|^2 - |\Psi|^2 \right)^2}
\label{Fmag}
\end{eqnarray}
which for large values of $\eta_{mag}$ effectively locks $|\Phi|$ to $|\Psi|$ (interpreting $|\Phi|=0$
 as the non-tilted phase and $|\Phi|=1$ as the tilted phase).  In addition, the tilted phases are known
 to exhibit direction-locking at various angles with respect to the bond orientation.
  This experimental information may be incorporated via an energy
\begin{eqnarray}
F_{dir} = \eta_{dir} \int d{\vec r} \ |\Phi|^2 |\Psi|^2 \left\{1 -  \cos{N\left( \phi-\theta-\theta_{dir} \right)} \right\},
\label{Fdir}
\end{eqnarray} 
where $\theta_{dir}$ is the equilibrium angle between 
the tilt direction and the crystal axis orientation.  

Note that the formulation just described is not meant to capture the
full physics of the phase transitions among the various tilted phases.
Previous workers have formulated Landau theories describing the 
phase transitions among the various condensed phases (those possessing
orientational order)\cite{Selinger,Kaganer}.  While one could attempt
to generalize this theory to describe transitions
among the various condensed phases (both tilted and non-tilted) and the
expanded phases (those with no orientational order and hence no ``crystal
axes"), we are not interested in transitions amongst all these phases.
The above formulation is only meant to capture the physics of the 
transition from a liquid-expanded 
(``LE") phase with $|\Psi|=0$ to a single {\em specific} tilted condensed phase
in which $|\Psi|=|\Phi|=1$.  
In the context of the literature surrounding tilted hexatic phases for
{\em achiral} molecules, 
$\theta_{dir}=0$ corresponds to an ``$L_2$''-like tilted phase 
(where the tilt points to the nearest neighboring molecule), 
$\theta_{dir}={\pi \over \ N}$ to an ``$L_2^*$''-like tilted phase 
(where the tilt points to the next nearest neighbor). Any other value of 
$\theta_{dir}$ is akin to an  ``$L_1^\prime$''-like phase~\cite{Fischer}
which has dynamic reflection-symmetry breaking. If the molecules are chiral,
there will be no chiral symmetry breaking transition and 
$\theta_{dir}$ will take some specific value.

Finally, we add an interaction between the tilt angle at the interface
and the interfacial orientation; this is done via the inclusion of the term
\begin{eqnarray}
F_{U\Phi} = \eta_{U\Phi} \int d{\vec r}\ { 1 \over 2} |\Phi| \left\{ e^{-i\beta}\Phi\partial_z U  + c.c. \right\}.
\label{Fuphi}
\end{eqnarray} 
In the sharp-interface limit, this interaction becomes a boundary condition on  the angle
of the tilt , $\phi$, at the interface.
To see this, consider a 
small, nearly planar portion of the interface.  Since $U$ varies most rapidly 
across the interface, we may approximate $U$ as a function of only the 
coordinate normal to the interface, $v \equiv  x\cos{\alpha} + y\sin{\alpha}$ 
(where $\alpha$ is the angle from the $x-axis$ to the gradient of $U$).  Then 
$\partial_z U = {1 \over 2} \left( \partial_x - i \partial_y \right) 
U \approx  { 1 \over 2}e^{-i\alpha} \partial_v U$, and $F_{U\Phi}$ becomes 
proportional to $\cos{\left(\phi - \beta - \alpha \right)}$.  So we expect 
the tilt to make an angle $\beta$ with respect to the interface normal at 
equilibrium.  For the growth process,  we expect this condition to be most strongly enforced 
at the dendrite tip as the diffusion gradient is enhanced there. Note that
a nonzero value of $\beta$ can only arise for achiral molecules.

We have used the phase-field model just described as a computational
stand-in for the ``true" sharp-interface model (as defined by
experimental observations).    In the limit of a sharp
interface, we expect the magnitudes of the order parameters
$|\Psi|$ and $|\Phi|$ to be locked to their equilibrium 
values.  That is, there is a region of liquid-expanded phase in which $|\Psi|=|\Phi|=0$
surrounding a region of tilted liquid-condensed (``solid") phase in which $|\Psi|=|\Phi|=1$, with a 
physically ``sharp" interface separating the two regions.
Meanwhile, the phase angles of the order parameters
(the bond direction $\theta$ and the tilt direction $\phi$) are free to vary continuously 
in the bulk ``solid".  The relationship between the tilt direction and the bond
direction has been observed experimentally, and is approximated here by 
the energies (\ref{Fmag}) and (\ref{Fdir}).  
While the boundary condition on the bond orientation
cannot (as yet) be observed experimentally, boundary conditions on the tilt at
the interface have been observed\cite{Fischer}.  We approximate these tilt
boundary conditions with the energy (\ref{Fuphi}).

 \section*{III.  Simulations}
We have performed numerical simulations of the above model
 (defined by equations (\ref{PsiEqn}),(\ref{UEqn}), and (\ref{PhiEqn}) ) with a free energy
\begin{eqnarray}
F[\Psi,\Phi,U] = \int d{\vec r} \ \kappa_{\Phi}^2 |\nabla \Phi|^2 + F_{\Psi}+ F_{mag} + F_{dir} + F_{U\Phi}
\label{Ftotal}
\end{eqnarray} 
where the surface tension due to $\Psi$ is the N-fold symmetric function given by equation
 (\ref{AnisotropicKappa}) and the surface tension due to $\Phi$ is isotropic.
For convenience we rescaled the model to have length units of $l_k\equiv D\Gamma/\kappa_0$ (the ``kinetics-limited diffusion length''), time units of $l_k^2/D$ (the related diffusion time), and an interface thickness of $\epsilon$ (see Ref. \cite{KupfermanNumerical} for details of rescalings).  Our present computing resources limited our simulations to 200x230 gridpoints (on a triangular lattice), which does not sufficiently resolve the interface.  We used a three-fold bond orientation order parameter ($N=3$) to help minimize competition between the implicit six-fold lattice anisotropy and the explicit anisotropy of the bond orientation (for further discussion, see Ref. \cite{Kam}).   

Clearly, we do not pretend to present quantitatively accurate simulations of these spiral growths.  Our goal here is to check that the set of interactions described above can indeed give rise to the spiral dendrites seen in the experiments.  Further progress would require more fully resolved simulations and systematic procedures for reducing the effects of lattice anisotropy.  We refer the concerned reader to the important work of Boesch, Mueller-Krumbhaar, and Shochet\cite{Boesch} which presents a detailed prescription for controlling the effects of lattice anisotropy in computer simulations of phase-field models.  We view such careful
studies as a necessary postlude to the phenomenological model presented here and as
a prerequisite for any quantitative comparisons with experiment.   

In general, the simulations 
below used $\epsilon = 0.4$, $\lambda = 10$,  $T = 0.6$, $N = 3$, 
$\Gamma_\Phi = 0.75 \Gamma$, 
$\kappa_{\Phi}\cong 10^{-4} \kappa_0$,$\eta_{mag} = 1.0$, and 
$\eta_{dir} = 1.0$.  Initial seeds were circles with radii of $6$
with uniform $\theta$-distributions whose three-fold directions are obtained by rotations of
 ${2 \pi \over \ 3}$ from the $+y$-axis; the initial $\phi$-distributions are
described below.  The undercooling at infinity was fixed at a value $\Delta$ 
(i.e. $U\left(\infty  \right)=-\Delta$) which varied in the simulations below.  
To prevent overcrowding in the figures, tilt vectors are shown for approximately one tenth of the points for which
 $|\Psi|\geq0.50$ (the ``solid"), resulting in domains which are sometimes 
slightly more planar than the fully sampled interfaces.  Bond 
orientations made an angle $\theta_{dir}$ with the 
tilt (to within a tenth of a degree) and do not appear in the figures.

The three-armed spiral dendrite of Figure 1a was achieved with a strong 
anisotropy (compared to the relatively slow dendrite growth speed, $\Delta=0.50$ and
$\eta_{\kappa}=2.0$), a chiral boundary condition on the tilt direction at the interface 
($\eta_{U\Phi}=2.5$, $\beta={-\pi \over \ 4}$) and a tilt locked parallel to the local crystal axis 
orientation ($\theta_{dir}=0$, analogous to the ``$L2$'' phase in the literature~\cite{Fischer}).
Within the first 100 timesteps, the initially radial $\phi$-distribution relaxed to
the domain structure shown in the inset of Figure 1, which is the three-fold analogy of the tilt 
domains seen experimentally in droplet textures of hexatic condensed phases~\cite{Fischer}.
In this case, the chirality of the material was expressed in the
boundary condition on the tilt which 
meant that the tilt rested at an angle of ${\pi \over 4}$ clockwise from the normal to the interface.  
This resulted in a clockwise bending of the crystal axes via the energy $F_{dir}$ and hence a 
clockwise turning of the growth direction.

As expected, there is a parameter regime in which the anisotropy is too weak (for a given 
dendrite growth speed) to support stable dendrites, resulting in a dense-branched structure.  The 
combination of a higher dendritic growth speed ($\Delta=0.90$) and a lower anisotropy strength 
($\eta_{\kappa}=0.25$) produced the pattern shown in Figure 1b (all 
other parameters are identical to those in Figure 1a). This pattern resembles
at least to some extent the chiral dense-branched structures 
seen in bacterial colony growth~\cite{bacteria}.

An interesting scenario is the case where the tilted phase is chiral
while the constituent molecules themselves are achiral.  
[For some discussion of chiral symmetry breaking by achiral molecules
in Langmuir monolayers and Langmuir-Blodgett films see References \cite{Selinger,Viswanathan}.]
This case is peculiar in that a given ``$L_1^\prime$"-like phase 
can produce either a right-handed or a left-handed spiral, depending
on the orientation of the domain walls of the early tilt field with
respect to the crystal axes.  An example of this is shown in Figure 2 
where stable dendrites ($\Delta=0.65$, $\eta_{\kappa}=2.0$) grow under the influence of an 
achiral tilt boundary condition ($\eta_{U\Phi}=2.5$, $\beta=0$) in an ``$L1^\prime$''-like tilted 
phase ($\theta_{dir}={-\pi \over \ 4}$).  Constant angular
displacements were added to otherwise radial $\phi$-distributions to
obtain the tilt domains (formed during the first 100 timesteps) shown
in the insets.  In Figure 2a, the tilt in the upper domain is locked ${\pi \over 4}$ radians counterclockwise from the crystal axis along the $+y$-axis.
  As the dendrite tip emerges along the $+y$-axis, the tilt rotates
clockwise to follow it.
To remain in the ``$L_1^\prime$"-like phase, the crystal axes (and hence,
the growth direction) of the added material also turn clockwise.
In Figure 2b, the tilt in the upper domain is locked ${\pi \over 4}$ 
counterclockwise from the crystal axis oriented ${\pi \over \ 6}$
below the $+x$-axis.  This time, as the dendrite tip emerges from the 
upper domain the tilt must rotate counterclockwise to follow it.
The growth direction will then turn counterclockwise as the crystal
axes rotate to remain in the existing ``$L_1^\prime$"-like phase.

\section*{IV.  Conclusion}
In summary, we have introduced a complex order parameter representing the tilt degree of 
freedom in an anisotropic complex-phase-field solidification model, and added terms to the free 
energy which represent interactions between the tilt, the solid-liquid interface, and the bond 
orientation.  Though these added terms have not been derived from thermodynamic identities, we appeal to the sharp interface model suggested by the experimental observations of tilted hexatic domains as a motivation for these terms.

We have shown that our model allows for the simulation of various chiral scenarios 
which produce spiral dendrites and chiral dense-branched structures.   We note in passing that 
we have had to allow for significant flexibility in the tilt-field and bond orientation in the bulk
of the material.  While this is reasonable for liquid crystal growths, presumably most other
crystalline systems (e.g. metallic) would be much more rigid and would probably preclude the 
possibility of spiral growth.
Nonetheless, this demonstration of a set of sufficient conditions for the
formation of these structures should help further the 
understanding of the intricate global morphologies seen in experiments
on diffusion-limited growth over the last decade.

\acknowledgements
We would like to acknowledge useful conversations with D. Kessler, O.
Schochet and E. Ben-Jacob. This work was supported in part by
NSF grant DMR94-15460.

\clearpage
\section*{Figure Captions}
\begin{itemize}
\item[FIG. 1.]  Growth shapes for chiral tilt boundary conditions ($\eta_{U\Phi}=2.5$, $\beta={-\pi 
\over \ 4}$) in an ``L2''-like tilted phase ($\theta_{dir}=0$) .
(a) Stable three-fold spiral dendrites for $\Delta=0.50$, $\eta_{\kappa}=2.0$.  (b)  Chiral 
dense-branching for $\Delta=0.90$ and $\eta_{\kappa}=0.25$.

\item[FIG. 2.]  Stable three-fold spiral dendrites for normal boundary condition on tilt 
($\eta_{U\Phi}=2.5$,$\beta=0$) in an ``$L1^\prime$''-like tilted phase ($\theta_{dir}={-\pi \over \ 
4}$), with $\Delta=0.65$, $\eta_{\kappa}=2.0$.  Early tilt domain formations are shown in the insets.

\end{itemize}

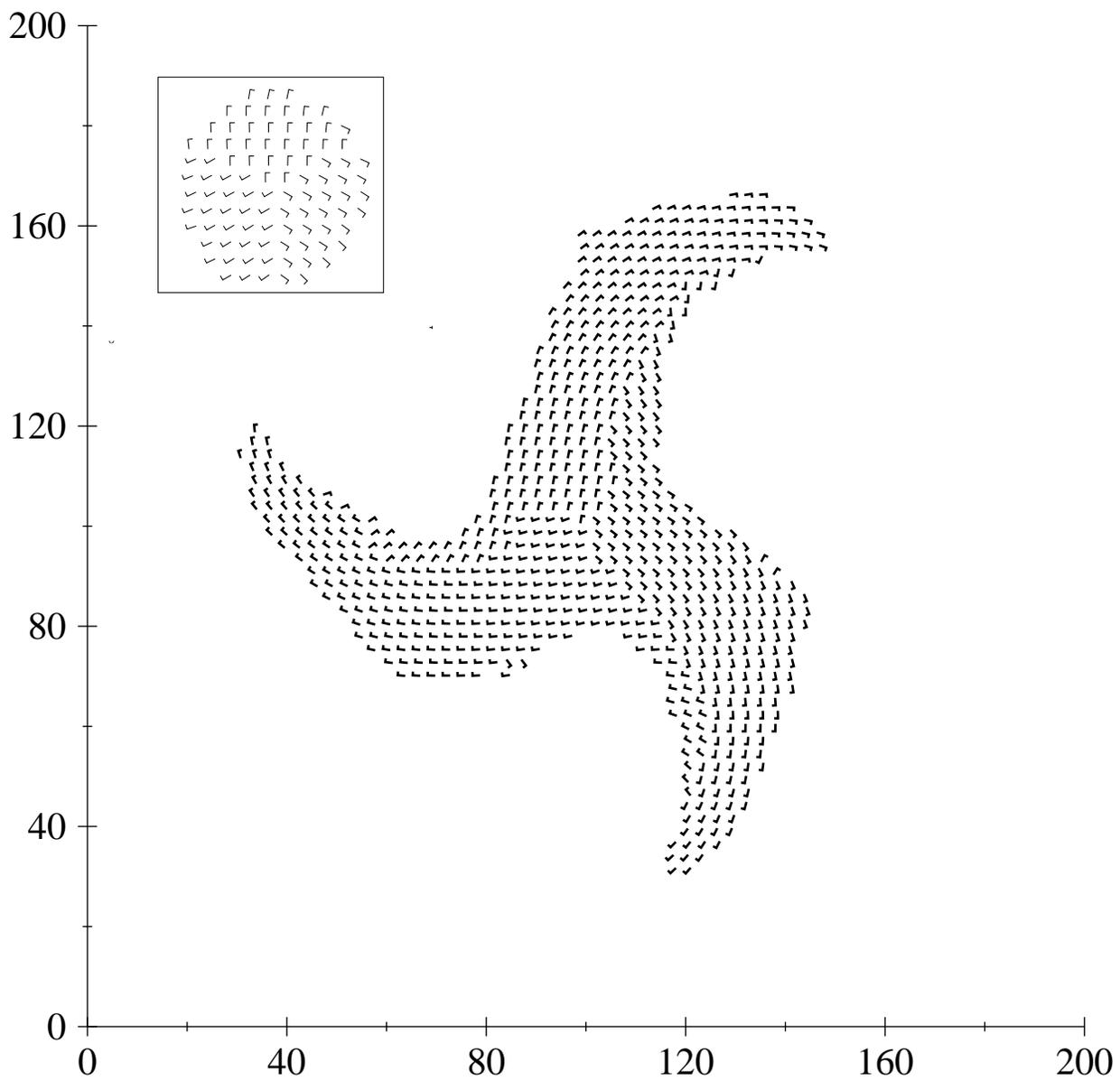
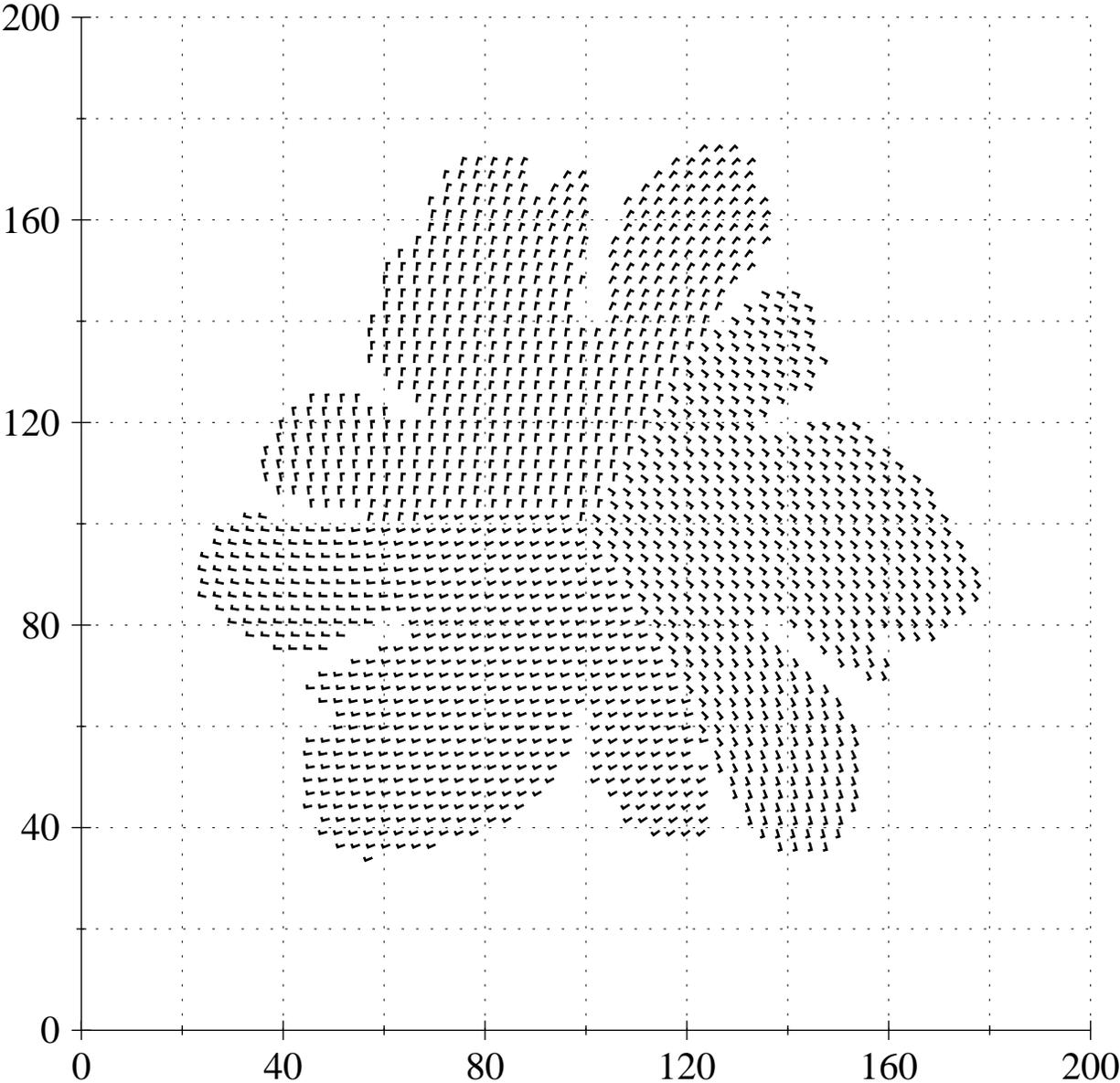
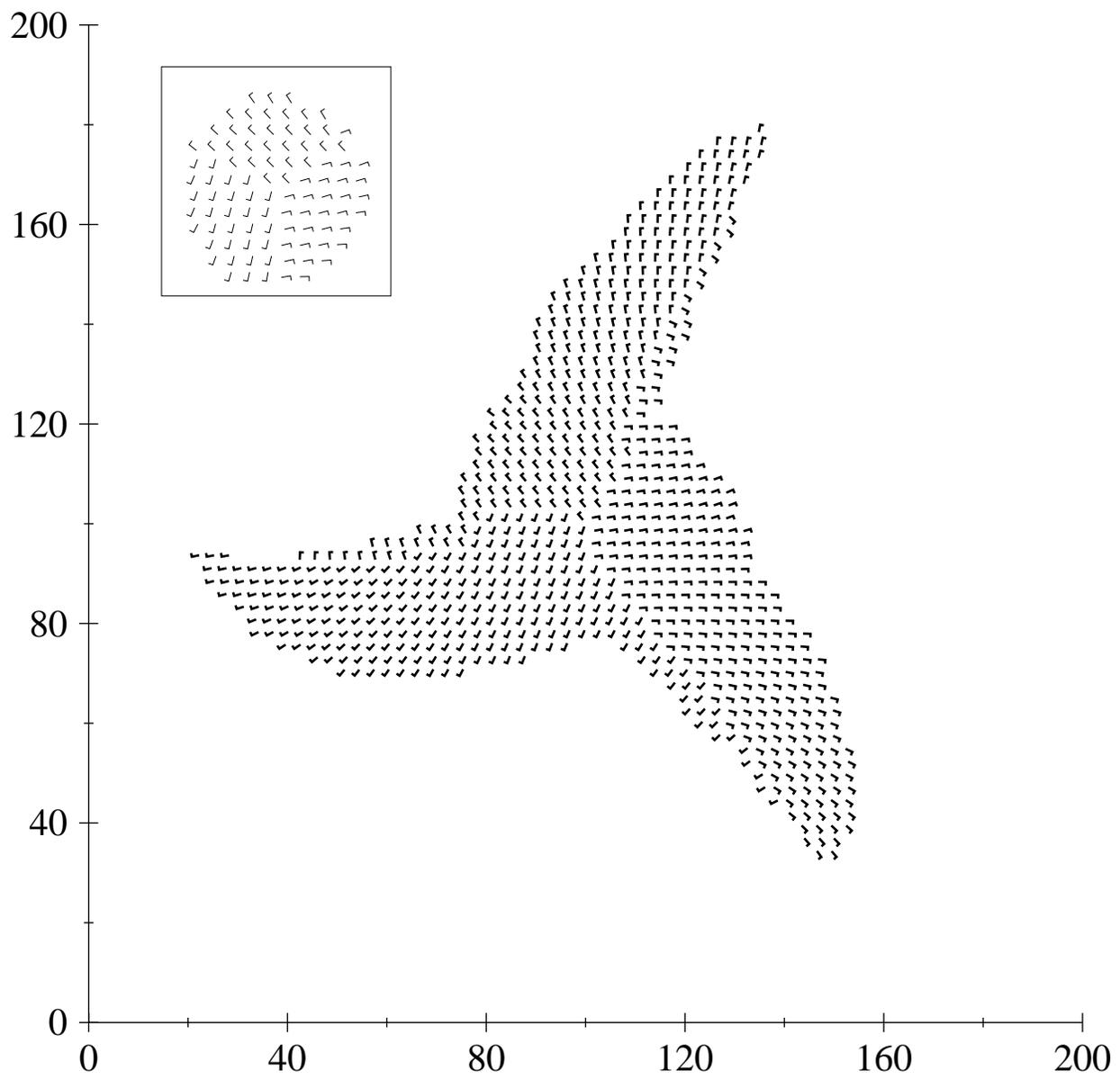
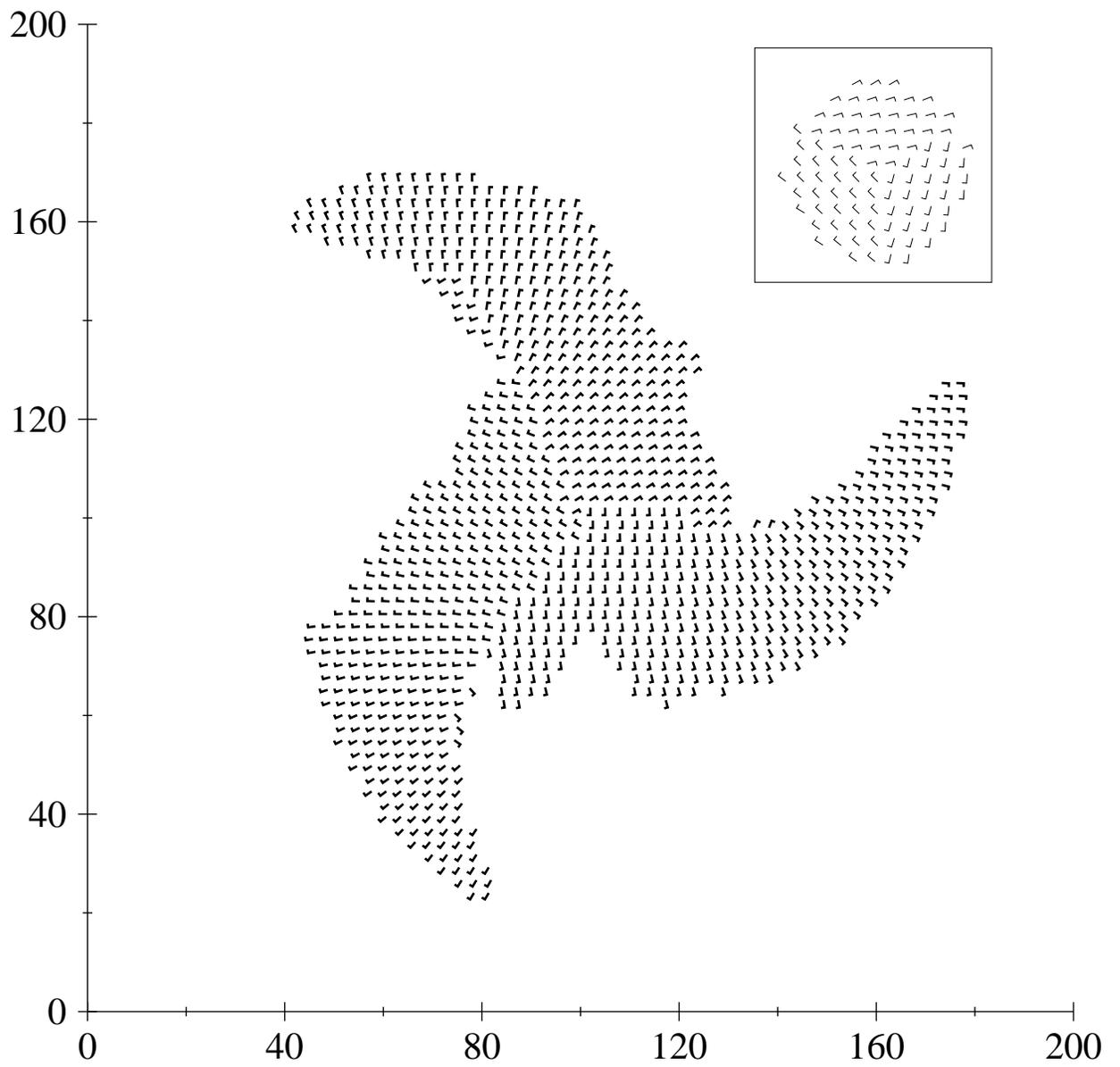
\end{document}